\documentclass[a4paper, amsmath, amssymb, 12pt]{article}

\usepackage[top=5.5cm, bottom=3.5cm, left=3.5cm, right=3.6cm]{geometry}
\usepackage[fleqn]{amsmath}
\usepackage{amssymb}
\usepackage{amsthm}

\begin{document}

\begin{center}
\large {\bf New properties of Special Functions and applications}

\vspace{0.4cm}

\large Vagner Jikia$^{1}$, Ilia Lomidze$^{2}$ \\
{\small $^{1}$ Javakhishvili State University, Tbilisi, Georgia, \\
$^{2}$ Georgian Technical University, Engineering Physics Department, Tbilisi, Georgia. \\ v\_jikia@yahoo.com, lomiltsu@gmail.com }
	
\end{center}
%
%
\vspace{0.2cm}

{\bf Abstract.} We derive some new results for the Mellin transform formulas as well as for the Gauss hypergeometric function. Also, we have found the relation between the Legendre functions of the second kind. We use some of these results in quantum mechanics of two charged particles of a continuous spectrum. \\
\\
{\bf Keywords and phrases:} Special functions, generalized functions, singular integrals. \\
\\
{\bf AMS subject classification (2010):} 33E99, 30G99 \\
\\
{\bf 1 Introduction} 
\\
\\
Special functions are important in different branches of theoretical physics and mathematics, for instance in particle physics [1]. That is why a great deal of investigations is devoted to this subject for so long time [2, 3].

We have determined the Beta function of the imaginary parameters [4]. The function we have defined is the singular generalized function and may be expressed through the delta function. Some applications of the method developed in [4] will be considered below. 

Using the aforementioned method we define the Mellin transform of the imaginary parameters for the function $ f(t) = 1,\;t \in \mathbb{R}. $ 

Also, we show that the Gauss hypergeometric function {\bf(HF)} 
\begin{eqnarray}  
\nonumber F\left( {\begin{array}{*{20}{c}}
a\\
c
\end{array};\,b,\,z} \right) = {}_2{F_1}\left( {a,\,b;\,c;\,z} \right) = {}_2{F_1}\left( {b,\,a;\,c;\,z} \right)
\end{eqnarray}
has the weak limit for $ z \to 1, $ for the set of the parameters $ a=2i\tau,\;\,b = i\tau,\;\,c = 2i\tau, $ $ \tau \in \mathbb{R}, $ nevertheless in this case $ {\mathop{\rm Re}\nolimits} \left( {c - a - b} \right) = 0 $ and so, the condition $ {\mathop{\rm Re}\nolimits} \left( {c - a - b} \right) > 0 $ is not fulfilled. Some of the consequences are considered below.\\
Finally, we consider the Legendre functions of the second kind and obtain the relation between $ 
Q_\nu\left( z \right)\ $ and $ 
Q_\nu ^\mu \left( z \right)\ $when $ {\mathop{\rm Re}\nolimits} \mu  = 0. $  Using the relation obtained, we derive some asymptotic formulas for the associated Legendre function $ 
Q_\nu ^{i\tau}\left(z\right),\;\; \tau\in\mathbb{R}.$ 
\\
\\
{\bf 2 Some special functions and the Dirac delta function} 
\\
\\
The Euler Beta function has many features which can be used to explain the physical property of strongly interacting particles [1]. This function is defined as follows [5]
\begin{eqnarray}
\,B(\alpha,\,\beta ) = \;\int\limits_0^1 {{t^{\alpha  - 1}}{{(1 - t)}^{\beta  - 1}}dt,} \\ [0.2em]
\nonumber{\mathop{\rm Re}\nolimits} \,(\alpha ) > 0,\,\,{\mathop{\rm Re}\nolimits} \,(\beta ) > 0.
\end{eqnarray}
Replacing the variable of integration in (1) by the formula
\begin{eqnarray}  
t = \frac{u}{{1 + u}}\quad \left( {u = \frac{t}{{1 - t}}} \right),
\end{eqnarray}
one obtains the representation:
\begin{eqnarray}  
\,B(\alpha ,\,\beta ) = \;\int\limits_0^\infty  {du} \frac{{{u^{\alpha  - 1}}}}{{{{\left( {1 + u} \right)}^{\alpha  + \beta }}}} .
\end{eqnarray}
The function that is determined by the expression (1) satisfies the Euler formula:
\begin{eqnarray}  
\,B(\alpha ,\,\beta ) = \frac{{\Gamma (\alpha )\Gamma (\beta )}}{{\Gamma (\alpha  + \beta )}},
\end{eqnarray}
where $ \Gamma (z) $  is the Euler gamma function (the Euler integral of the second kind):
\begin{eqnarray}  
& \nonumber \Gamma (z) = \int\limits_0^\infty  {\,{t^{z - 1}}\exp ( - t)dt} , \\ [0.2em]
& \nonumber{\mathop{\rm Re}\nolimits} z > 0.
\end{eqnarray}
The beta function is an analytic function in the domain of its definition.

Taking into account wide application in physics, we study an associated Legendre function of the second kind (see, e. g., [12], N 8.711(4), p. 960). 

We consider the integral representation
\begin{flalign}
& \nonumber Q_\nu ^\mu \left( z \right) = \exp (i\pi \mu )\dfrac{{\Gamma \left( {\nu  + 1} \right)}}{{\Gamma \left( {\nu  - \mu  + 1} \right)}} \\ 
& \;\;\;\;\;\;\;\;\;\;\;\;\;\;\;\;\;\;\;\;\;\;\;\;\;\;\;\;\;\;\;\; \times \int\limits_0^\infty  {dt\cosh \left( {\mu t} \right){{\left( {z + \sqrt {{z^2} - 1} \cosh \left( t \right)\,} \right)}^{ - \nu  - 1}}} , \\ 
& \;\;\;\;\;\;\;\;\;\; \nonumber {\mathop{\rm Re}\nolimits} \left( {\nu  + \mu } \right) >  - 1,\;\;\,\nu  \ne  - 1,\; - 2,\;.\;.\;.\;,\;\;\; \left| {\arg \,(z \pm 1)} \right| < \pi .
\end{flalign}
For $ \mu  = 0 $ one obtains from (5) the Legendre function of the second kind:
\begin{eqnarray} 
{Q_\nu }\left( z \right) = \int\limits_0^\infty  {dt{{\left( {z + \sqrt {{z^2} - 1} \cos h\left( t \right)} \right)}^{ - \nu  - 1}}} .
\end{eqnarray}
Below we find and investigate the relation between the functions (6) and (5) for $ {\mathop{\rm Re}\nolimits} \mu  = 0. $

For completeness, we recall certain properties of the Dirac delta function.

According to the definition of the Dirac function (see e.g. [6], p. 99) we assume that it is defined as the weak limit of a sequence of some approximate functions $ {\omega _\varepsilon }(x): $ 
\begin{eqnarray} 
\delta (x)\mathop  = \limits^{weak} \mathop {\lim }\limits_{\varepsilon  \to 0 + } {\omega _\varepsilon }(x),\;\;x \in \mathbb{R}, 
\end{eqnarray}
where the approximate function $ {\omega _\varepsilon }(x) $ can be constructed as (see [6], pp. 86-87)
\begin{eqnarray} 
{\omega _\varepsilon }(x) = {\varepsilon ^{ - 1}}\eta ({\varepsilon ^{ - 1}}x),
\end{eqnarray} 
for any bounded finite function $ \eta (x), $ such that 
\begin{eqnarray} 
\int\limits_{ - \infty }^{ + \infty } {\eta (x)dx = 1.}
\end{eqnarray} 
Below we shall use function $ \eta (x) = {[\pi (1 + {x^2})]^{ - 1}}. $ 

Another mostly symbolic but rather common definition (see e.g. [6], pp.84-85) considers the Dirac delta as singular generalized function which satisfies the (weak) equality 
\begin{equation}
\int\limits_a^b {\varphi (x)\delta (x)dx} = \begin{cases}
&\varphi (0), \quad  0 \in (a,b),\\
&\varphi (0)/2, \quad a = 0,\;or\;b = 0, \\
&{0,  \quad 0 \notin [a,b]}
\end{cases}
\end{equation}
for any continuous function $ \varphi (x) $ defined on the interval $ [a,b]. $

We study the behavior of the $ B(\alpha ,\beta ) $ function and have shown [4] that for $ \beta = - \alpha = i\tau $ the following relations are valid:
\begin{eqnarray} 
\nonumber & B(i\tau,\, - i\tau ) = \mathop {\lim }\limits_{\varepsilon  \to 0 + } B(\varepsilon  + i\tau ,\,\varepsilon - i\tau ) = \\ [0.2em]
& = \mathop {\lim }\limits_{\varepsilon  \to 0 + } \,\int\limits_0^1 {{t^{\varepsilon  + i\tau  - 1}}} {\left( {1 - t} \right)^{\varepsilon  - i\tau  - 1}}dt = \int\limits_0^1 {{t^{i\tau  - 1}}} {\left( {1 - t} \right)^{ - i\tau  - 1}}dt = 2\pi \delta \left( \tau \right). 
\end{eqnarray}
So, the function $ B(i\,\tau ,\, - i\,\tau ) $ which we have defined is the singular generalized function and may be presented in the explicit form:
\begin{eqnarray} 
B(i\,\tau ,\, - i\,\tau ) = \int\limits_0^1 {{t^{i\tau  - 1}}} {\left( {1 - t} \right)^{ - i\tau  - 1}}dt = 2\pi \delta \left( \tau \right).
\end{eqnarray}
\\
\\
{\bf 3 Applications} \\

{\bf The Mellin transform.} Using the substitution (2) in the formula (11) one obtains the representation $ \left( {\tau  \in \mathbb{R} } \right) $:
\begin{eqnarray} 
B(i\,\tau ,\, - i\,\tau ) = \mathop {\lim }\limits_{\varepsilon  \to 0 + } \,\int\limits_0^\infty  {dt} \frac{{{t^{\varepsilon  + i\tau  - 1}}}}{{{{\left( {1 - t} \right)}^{2\varepsilon }}}} = \int\limits_0^\infty  {dy\,{y^{i\tau  - 1}}}  = 2\pi \delta (\tau ).
\end{eqnarray}
Consider the Mellin integral [7]:
\begin{eqnarray} 
\phi (z) = \int\limits_0^\infty  {{t^{z - 1}}} f(t)dt,\;\;\,z \in \mathbb{C}, 
\end{eqnarray} 
where $ {t^{z - 1}}f(t) \in L\left( {0,\infty } \right) $ and the function $ f(t) $ is bounded for arbitrary real $ t. $ Under this conditions, $ \phi (z) $ is continuous and $ f(t) $ can be expressed in terms of the inverse transformation [7]:
\begin{eqnarray} 
f\left( t \right) = \frac{1}{{2\pi i}}\int\limits_{ - i\infty }^{ + i\infty } {{t^{ - z}}} \phi \left( z \right)dz.
\end{eqnarray} 
By (14) and (13) the integral:	
\begin{eqnarray} 
\int\limits_0^\infty  {dy\,{y^{ix - 1}}} = 2\pi \delta \left( x \right)
\end{eqnarray}
is the Mellin integral of the function $ f\left( t \right) = 1 $ for $ 
{\mathop{\rm Re}\nolimits} \left( z \right) = 0. $ Using the expression (15) and the function $ 2\pi \delta \left( x \right) $ one can define the inverse transformation:
\begin{eqnarray} 
f\left( t \right) = \frac{1}{{2\pi i}}\int\limits_{ - i\infty }^{ + i\infty } {d\left( {ix} \right){t^{ - ix}}2\pi \delta \left( x \right) = 1} .
\end{eqnarray}

Thus, for $ {\mathop{\rm Re}\nolimits} \left( z \right) = 0 $ Mellin's formulas for the function $ f\left( t \right) = 1 $ are reasonable in the sense of singular generalized functions. The expression (16) shows that the function which is obtained by the Mellin transform of the function $ f\left( t \right) = 1 $ is not continuous. In addition, note that the Mellin transform of the function $ f\left( t \right) = 1 $ coincides in this case with its Fourier transform. Similar results for the case $ {\mathop{\rm Re}\nolimits} \left( z \right) \ne 0 $ holds [8].

{\bf The Gauss Hypergeometric Function.} Consider Gauss well known formula (see e. g. [5], 2.1.3(14), p. 73):
\begin{eqnarray} 
&F\left( {\begin{array}{*{20}{c}}
a\\
c
\end{array};\,b,\,1} \right) = \dfrac{{\Gamma \left( c \right)\Gamma \left( {c - a - b} \right)}}{{\Gamma \left( {c - a} \right)\Gamma \left( {c - b} \right)}}, \\ [0.2em]
&\nonumber {\mathop{\rm Re}\nolimits} \left( c \right) > {\mathop{\rm Re}\nolimits} \left(b \right) > 0,\quad {\mathop{\rm Re}\nolimits} \left( {c - a - b} \right) > 0.
\end{eqnarray}

For the set of parameters $ a = 2i\tau ,\;\,b = \varepsilon  + i\tau ,\;\,c = 2\varepsilon  + 2i\tau\; 
\left( {\varepsilon  > 0} \right) $ one can write:
\begin{eqnarray} 
F\left( {\begin{array}{*{20}{c}}
{2i\tau }\\
{2\varepsilon  + 2i\tau }
\end{array};\,\varepsilon  + i\tau ,\,1\,} \right) = \frac{{\Gamma \left( {2\varepsilon  + 2i\tau } \right)\Gamma \left( {\varepsilon  - i\tau } \right)}}{{\Gamma \left( {2\varepsilon } \right)\Gamma \left( {\varepsilon  + i\tau } \right)}}.
\end{eqnarray}

Using the method proposed in [4], let us calculate the weak limit:
\begin{eqnarray} 
\mathop {\lim }\limits_{\varepsilon  \to 0 + } F\left( {\begin{array}{*{20}{c}}
{2i\tau }\\
{2\varepsilon  + 2i\tau }
\end{array};\,\varepsilon  + i\tau ,\,1\,} \right).
\end{eqnarray}
By the Legendre formula
\begin{eqnarray} 
\nonumber \Gamma \left( {2z} \right) = \frac{{{2^{2z - 1}}}}{{\sqrt \pi  }}\Gamma \left( z \right)\Gamma \left( {z + \frac{1}{2}} \right)
\end{eqnarray}
after simplifications, one gets:
\begin{eqnarray} 
F\left( {\begin{array}{*{20}{c}}
{2i\tau }\\
{2\varepsilon  + 2i\tau }
\end{array};\,\varepsilon  + i\tau ,\,1\,} \right) = \frac{{{2^{2\left( {\varepsilon  + i\tau } \right) - 1}}}}{{\sqrt \pi  }}\frac{{\Gamma \left( {\varepsilon  + i\tau  + {1 \mathord{\left/
 {\vphantom {1 2}} \right.
 \kern-\nulldelimiterspace} 2}} \right)\Gamma \left( {\varepsilon  - i\tau } \right)}}{{\Gamma \left( {2\varepsilon } \right)}}.
\end{eqnarray}
Using known property of the gamma function $ z\Gamma (z) = \Gamma (1 + z) $, formula (21) can be rewritten as follows:	 
\begin{eqnarray} 
F\left( {\begin{array}{*{20}{c}}
{2i\tau }\\
{2\varepsilon  + 2i\tau }
\end{array};\,{\varepsilon  + i\tau,\,1\,}} \right) = f\left({\varepsilon,\;\tau}\right)\left(\varepsilon + i\tau\right) {\omega _\varepsilon}\left(\tau\right),
\end{eqnarray}
where $ \; {\omega _\varepsilon }(\tau ) = {\pi ^{ - 1}}\varepsilon {\left( {{\varepsilon ^2} + {\tau ^2}} \right)^{ - 1}} $ is defined according to (8), and
\begin{eqnarray} 
f\left( {\varepsilon ,\;\tau } \right) = \sqrt \pi  {2^{2\left( {\varepsilon  + i\tau } \right)}}\frac{{\Gamma \left( {\varepsilon  + i\tau  + {1 \mathord{\left/
 {\vphantom {1 2}} \right.
 \kern-\nulldelimiterspace} 2}} \right)\Gamma \left( {\varepsilon  - i\tau  + 1} \right)}}{{\Gamma \left( {2\varepsilon  + 1} \right)}}.
\end{eqnarray}
The function $ f\left( {\varepsilon ,\;\tau } \right) $ is an analytic at the point $ \varepsilon  = 0, $ 		$ \tau  = 0 $ of the complex plane and $ f(\varepsilon ,\tau ) \in \mathbb{C^\infty }. $ So, it can be expanded in the neighborhood of the point $ \left( {0,\;\tau } \right), $  $ \tau\in\mathbb{R}. $ According to the Taylor formula with remainder in the Lagrange form 	
\begin{eqnarray} 
\nonumber f(\varepsilon ,\;\tau ) = f(0,\;\tau ) + \frac{1}{{1!}}{f'_\varepsilon }(0,\;\tau )\varepsilon  + \frac{1}{{2!}}{f''_{\varepsilon \varepsilon }}(\xi ,\;\tau ){\varepsilon ^2},\quad 0 < \xi  < \varepsilon ,\quad \tau  \in \mathbb{R},
\end{eqnarray}
one obtains
\begin{flalign}
& \nonumber F\left( {\begin{array}{*{20}{c}}
{2i\tau}\\
{2\varepsilon + 2i\tau}
\end{array};\,\varepsilon + i\tau,\,1\,}\right)\\ 
&\;\;\;\;\;\;\;
=\left(\varepsilon + i\tau\right)\left( {f(0,\;\tau ) + \dfrac{1}{{1!}}{{f'}_\varepsilon }(0,\;\tau )\varepsilon  + \dfrac{1}{{2!}}{{f''}_{\varepsilon \varepsilon }}(\xi,\;\tau ){\varepsilon ^2}} \right){\omega _\varepsilon}\left(\tau\right),\\  
& \;\;\;\;\;\;\;\;\;\;\;\;\;\;\;\;\;\;\;\;\;\;\;\;\;\;\;\;\;\;\;\;\;\;\;\;\;\;\;\;\;\;\; \nonumber 0 < \xi  < \varepsilon ,\quad \tau  \in \mathbb{R},
\end{flalign}
where 
\begin{eqnarray} 
&\nonumber f\left( {0,\,\tau } \right) = \frac{{{4^{i\tau }}}}{{\sqrt \pi  }}\Gamma \left( {i\tau  + {1 \mathord{\left/
{\vphantom {1 2}} \right.
\kern-\nulldelimiterspace} 2}} \right)\Gamma \left( { - i\tau  + 1} \right), \;\;\;\;\;\;\;\;\;\;\;\;\;\;\;\;\;\;\;\;\;\;\;\;\;\;\;\;\;\;\;\;\;\;\;\;\;\;\;\;\;\;\;\;\;\;\;\;\;\;\;\\ 
&\nonumber{f'_\varepsilon }\left( {0,\,\tau } \right) = f\left( {0,\;\tau } \right)\left[ {\psi \left( {i\tau  + {1 \mathord{\left/
{\vphantom {1 2}} \right.
\kern-\nulldelimiterspace} 2}} \right) + \psi \left( { - i\tau  + 1} \right) + 2\gamma  + \ln 4} \right],\;\;\;\;\;\;\;\;\;\;\;\;\;\;\;\;\;\;\; \\
&{f''_{\varepsilon \varepsilon }}\left( {\xi ,\,\tau } \right) = f\left( {\xi ,\;\tau } \right)\left\{ {\psi '\left( {\xi  + i\tau  + {1 \mathord{\left/
 {\vphantom {1 2}} \right.
 \kern-\nulldelimiterspace} 2}} \right) + \psi '\left( {\xi  - i\tau  + 1} \right) - 4\psi '\left( {2\xi  + 1} \right)} \right. \;\;\;\;\\
&\nonumber \;\;\;\;\;\;\;\;\;\;+ \left. {{{\left[ {\psi \left( {\xi  + i\tau  + {1 \mathord{\left/
 {\vphantom {1 2}} \right.
 \kern-\nulldelimiterspace} 2}} \right) + \psi \left( {\xi  - i\tau  + 1} \right) - 2\psi \left( {2\xi  + 1} \right) + \ln 4} \right]}^2}} \right\}, \\
&\;\;\;\;\;\;\;\;\;\;\;\;\;\;\;\nonumber 0 < \xi < \varepsilon ,\quad \tau  \in \mathbb{R},
\end{eqnarray}
and $ \psi \left( x \right) $ is the digamma function: 
\begin{eqnarray} 
\nonumber \psi \left( x \right) = \frac{d}{{dx}}\Gamma \left( x \right) = {{\Gamma '\left( x \right)} \mathord{\left/
 {\vphantom {{\Gamma '\left( x \right)} {\Gamma \left( x \right)}}} \right.
 \kern-\nulldelimiterspace} {\Gamma \left( x \right)}}.
\end{eqnarray}
Therefore
\begin{eqnarray} 
\nonumber & f\left( {0,\,0} \right) = 1,\quad {f'_\varepsilon }\left( {0,\,0} \right) = 0,\quad {f''_{\varepsilon \varepsilon }}(\xi ,\;\tau ) \le M < \infty. 
\end{eqnarray} 
By definition (7), one can write:
\begin{eqnarray} 
\delta (\tau )\mathop  = \limits^{weak} \mathop {\lim }\limits_{\varepsilon  \to 0 + } {\omega _\varepsilon }(\tau ) = \mathop {\lim }\limits_{\varepsilon  \to 0 + } \frac{1}{\pi }\frac{\varepsilon }{{{\varepsilon ^2} + {\tau ^2}}},
\end{eqnarray}	
which has the following integral form:
\begin{flalign} 
&\nonumber \int\limits_a^b {d\tau \varphi (\tau )\delta (\tau )} = \\
&\;\;\;\;\;=\mathop {\lim }\limits_{\varepsilon  \to 0 + } \int\limits_a^b {d\tau \varphi (\tau )\frac{1}{\pi }\frac{\varepsilon }{{{\varepsilon ^2} + {\tau ^2}}}}  =\begin{cases}
&\varphi (0), \quad  0 \in (a,b),\\
&\varphi (0)/2, \quad a = 0,\;or\;b = 0, \\
&{0,  \quad 0 \notin [a,b]}
\end{cases}
\end{flalign}
where $ {\varphi (\tau )} $ is a continuous and bounded function. According to the formula (27), from the equality (24) one obtains:
\begin{eqnarray} 
\mathop {\lim }\limits_{\varepsilon  \to 0 + } \int\limits_a^b {d\tau \varphi (\tau )F\left( {\begin{array}{*{20}{c}}
{2i\tau }\\
{2\varepsilon  + 2i\tau }
\end{array};\,\varepsilon  + i\tau ,\,1\,} \right)}  = 0.
\end{eqnarray}
By the definition of the weak limit (see, e. g. [10], p. 353), one can write:
\begin{eqnarray} 
\mathop {\lim }\limits_{\varepsilon  \to 0 + } F\left( {\begin{array}{*{20}{c}}
{2i\tau }\\
{2\varepsilon  + 2i\tau }
\end{array};\,\varepsilon  + i\tau ,\,1\,} \right)\mathop  = \limits^{weak} 0.
\end{eqnarray}

We define $ F\left( {\begin{array}{*{20}{c}} {2i\tau }\\
{2i\tau }
\end{array};\,i\tau ,\,1\,} \right) $  as follows:
\begin{flalign} \label{sixty}
& \nonumber \;\;\;\; F\left( {\begin{array}{*{20}{c}}
{2i\tau }\\
{2i\tau }
\end{array};\,i\tau ,\,1\,} \right)\mathop = \limits^{weak} \mathop {\lim }\limits_{\varepsilon \to 0 + } F\left( {\begin{array}{*{20}{c}}
{2i\tau } \\ [0.4em]
{2\varepsilon  + 2i\tau }
\end{array};\,\varepsilon + i\tau ,\,1\,} \right) = \\ 
& \;\;\;\;\;\;\;\;\;\;\;\;\;\;\;\;\;\;\;\;\;\;\;\;\;\;\;\;\;\;\;\;\;\;\;\;\;\;\;\;\;\;\;\; = \mathop {\lim }\limits_{\varepsilon  \to 0 + } F\left( {\begin{array}{*{20}{c}}
{\varepsilon  + i\tau }\\
{2\varepsilon  + 2i\tau }
\end{array};2i\tau \,,\,1\,} \right).
\end{flalign}
Thus, even if the conditions $ {\mathop{\rm Re}\nolimits} \left( c \right) > {\mathop{\rm Re}\nolimits} \left( b \right) > 0,\;\;{\mathop{\rm Re}\nolimits} \left( {c - a - b} \right) > 0 $ are not fulfilled (see 18) there exists the weak limit:
\begin{eqnarray} 
F\left( {\begin{array}{*{20}{c}}
{2i\tau }\\
{2i\tau }
\end{array};\,i\tau ,\,1\,} \right)\mathop  = \limits^{weak} 0.
\end{eqnarray}

{\bf Corollary.} Using the well-known formula ([5], 2.8(4), p.109)
\begin{eqnarray} 
F\left( {\begin{array}{*{20}{c}}
a\\
b
\end{array};\,b,\,z} \right) = {\left( {1 - z} \right)^{ - a}},\quad \left| z \right| < 1
\end{eqnarray}
and the equality (31), one obtains:
\begin{eqnarray} 
\mathop {\lim }\limits_{z \to 1} F\left( {\begin{array}{*{20}{c}}
{i\tau }\\
{2i\tau }
\end{array};\,2i\tau ,\,z} \right) = F\left( {\begin{array}{*{20}{c}}
{2i\tau }\\
{2i\tau }
\end{array};\,i\tau ,\,1\,} \right) = \mathop {\lim }\limits_{z \to 1} {\left( {1 - z} \right)^{ - i\tau }}\mathop  = \limits^{weak} 0.
\end{eqnarray}
By relation (32), the limit (33) is determined in the area $ \arg (1 - z) = \xi  \in ({{ - \pi } \mathord{\left/
 {\vphantom {{ - \pi } 2}} \right.
 \kern-\nulldelimiterspace} 2},{\pi  \mathord{\left/
 {\vphantom {\pi  2}} \right.
 \kern-\nulldelimiterspace} 2}). $
 
Taking into account
\begin{eqnarray} 
\nonumber \mathop {\lim }\limits_{z \to 1} {(1 - z)^{ - i\tau }} = \exp \left( {\xi \tau } \right) \times \mathop {\lim }\limits_{z \to 1} {\left| {1 - z} \right|^{ - i\tau }}\mathop  = \limits^{weak} 0, \quad \tau  \in \mathbb{R},
\end{eqnarray}
one gets:
\begin{eqnarray} 
\mathop {\lim }\limits_{z \to 1} {\left| {1 - z} \right|^{ - i\tau }} = \mathop {\lim }\limits_{z \to 1} \exp \left( { - i\tau \ln \left| {1 - z} \right|} \right)\mathop  = \limits^{weak} 0.
\end{eqnarray}
From (34) one obtains:
\begin{eqnarray} 
\mathop {\lim }\limits_{z \to 1} \cos \left( {\tau \ln \left| {1 - z} \right|} \right)\mathop  = \limits^{weak} 0,\,\,\,\,\,\mathop {\lim }\limits_{z \to 1} \sin \left( {\tau \ln \left| {1 - z} \right|} \right)\mathop  = \limits^{weak} 0.
\end{eqnarray}
As far as $ \quad \left| z \right| < 1, $ one can write:
\begin{eqnarray} 
\mathop {\lim }\limits_{z \to 1} \cos \left[ {\tau \ln (1 - z)} \right]\mathop  = \limits^{weak} 0,\,\,\,\,\,\mathop {\lim }\limits_{z \to 1} \sin \left[ {\tau \ln (1 - z)} \right]\mathop  = \limits^{weak} 0.
\end{eqnarray}

Similarly, one obtains:
\begin{eqnarray} 
\nonumber & {\left( {\dfrac{{1 - z}}{{1 + z}}} \right)^{ - i\tau }} = \exp \left[ { - i\tau \ln \left( {\dfrac{{1 - z}}{{1 + z}}} \right)} \right] = \\ [0.2em]
& = \cos \left[ {\tau \ln \left( {1 - z} \right) - \tau \ln \left( {1 + z} \right)} \right] - i\sin \left[ {\tau \ln \left( {1 - z} \right) - \tau \ln \left( {1 + z} \right)} \right].
\end{eqnarray}
According to (36) and (37), one can write:
\begin{eqnarray} 
\nonumber \mathop {\lim }\limits_{z \to 1} {\left( {\frac{{1 - z}}{{1 + z}}} \right)^{ - i\tau }} = \mathop {\lim }\limits_{z \to 1} \cos \left[ {\tau \ln (1 - z)} \right] - i\mathop {\lim }\limits_{z \to 1} \sin \left[ {\tau \ln (1 - z)} \right]\mathop  = \limits^{weak} 0.
\end{eqnarray}
In addition, from the equalities (36) we obtain:
\begin{eqnarray} 
\nonumber \mathop {\lim }\limits_{t \to 0} {t^{i\tau }}\mathop  = \limits^{weak} 0.
\end{eqnarray}

{\bf The Legendre Functions.} Now let us investigate the relation between the functions (6) and (5) for $ {\mathop{\rm Re}\nolimits} \mu  = 0. $ In this case, one obtains from (5):
\begin{flalign}
& \nonumber \;\; Q_\nu ^{i\tau }\left( a \right) = \exp ( - \pi \tau )\frac{{\Gamma \left( {\nu  + 1} \right)}}{{\Gamma \left( {\nu  - i\tau  + 1} \right)}} \\
& \;\;\;\;\;\;\;\;\;\;\;\;\;\;\;\;\;\;\;\;\;\;\;\;\;\;\;\;\;\;\;\; \times \int\limits_0^\infty{dt\cos \left( {\tau t} \right){{\left( {z + \sqrt {{z^2} - 1} \cosh \left( t \right)\,} \right)}^{ - \nu  - 1}}} , \\ 
\nonumber & \;\;\;\;\;\;\;\;\;\;\;\;\;\;\;\;\;\;\;\;\;\;\;\; {\tau \in \mathbb{R}},\quad {\mathop{\rm Re}\nolimits} \left( \nu  \right) >  - 1,\quad \left| {\arg \left( {z \pm 1} \right)} \right| < \pi .
\end{flalign}
If $ z = a, $ where $ a $ is a real number greater than 1, the following relation holds:  
\begin{eqnarray} 
{\left( {a + \sqrt {{a^2} - 1} \cosh \left( t \right)} \right)^{ - \nu  - 1}} = 0,\quad t \to \infty ,\quad {\mathop{\rm Re}\nolimits} \left( \nu  \right) >  - 1.
\end{eqnarray}
Due to (39) one gets from (38):
\begin{flalign}
\nonumber & Q_\nu ^{i\tau }\left( a \right) = \exp ( - \pi \tau )\frac{{\Gamma \left( {\nu  + 1} \right)}}{{\Gamma \left( {\nu  - i\tau  + 1} \right)}} \\
& \;\;\;\;\;\;\;\;\;\;\;\;\;\;\;\;\; \times \int\limits_0^A {dt\cos \left( {\tau t} \right){{\left( {a + \sqrt {{a^2} - 1} \cosh \left( t \right)\,} \right)}^{ - \nu  - 1}}}  + O\left( A \right),
\end{flalign}
where $ A < \infty . $ Similarly, we have:
\begin{eqnarray} 
{Q_\nu }\left( a \right) = \int\limits_0^A {dt{{\left( {a + \sqrt {{a^2} - 1} \cos h\left( t \right)} \right)}^{ - \nu  - 1}}}  + O\left( A \right).
\end{eqnarray}  
Note also that the function (39) is the impulse function and	
\begin{eqnarray} 
\nonumber {\left( {a + \sqrt {{a^2} - 1} \cosh \left( t \right)} \right)^{ - \nu  - 1}} \ge 0,\quad a > 1.
\end{eqnarray}
Thus, in the integral formula (40) one can write:	
\begin{flalign}
\nonumber & Q_\nu ^{i\tau }\left( a \right) = \exp ( - \pi \tau )\frac{{\Gamma \left( {\nu  + 1} \right)}}{{\Gamma \left( {\nu  - i\tau  + 1} \right)}}\cos \left( {\tau \eta } \right) \\
& \;\;\;\;\;\;\;\;\;\;\;\;\;\;\;\;\;\;\;\;\;\;\;\;\;\;\;\;\; \times \int\limits_0^A {dt{{\left( {a + \sqrt {{a^2} - 1} \cosh \left( t \right)\,} \right)}^{ - \nu  - 1}}}  + O\left( A \right), \\
\nonumber & \;\;\;\;\;\;\;\;\;\;\;\;\;\;\;\;\;\;\;\;\;\;\;\;\;\;\; 0 \le \eta  \le A,\quad {\mathop{\rm Re}\nolimits} \nu  >  - 1.
\end{flalign}
Substituting (41) into (42) one gets:
\begin{eqnarray} 
& Q_\nu ^{i\tau }\left( a \right) = \exp ( - \pi \tau )\dfrac{{\Gamma \left( {\nu  + 1} \right)}}{{\Gamma \left( {\nu  - i\tau  + 1} \right)}}\cos \left( {\tau \eta } \right){Q_\nu }\left( a \right),\quad a > 1, \\ [0.2em]
\nonumber & {\mathop{\rm Re}\nolimits} \nu  >  - 1,\quad {\eta, \; \tau \in \mathbb{R}}.
\end{eqnarray}

The function $ Q_\nu ^\mu \left( z \right) $ is single-valued and regular in the complex plane having a cut along the interval $ \left( { - \infty ,\,1} \right] $ (see e. g. [12], p. 959). Besides, the function $ {Q_\nu }\left( z \right) $ is single-valued and analytic in the same region (see e. g. [13], p. 169). Thus, by analytic continuation, one can write:
\begin{eqnarray} 
& Q_\nu ^{i\tau }\left( z \right) = \exp ( - \pi \tau )\dfrac{{\Gamma \left( {\nu  + 1} \right)}}{{\Gamma \left( {\nu  - i\tau  + 1} \right)}}\cos \left( {\tau \eta } \right){Q_\nu }\left( z \right), \\ [0.2em]
\nonumber & {\eta, \; \tau \in \mathbb{R}},\quad {\mathop{\rm Re}\nolimits} \nu  >  - 1,\quad \left| {\arg \left( {z - 1} \right)} \right| < \pi .
\end{eqnarray}

To define the real parameter $ \eta, $ let us rewrite (44) in the region $ \left| z \right| \to \infty . $ Inserting the asymptotic relation (see e. g. [5], 3.9.2(21), p.165)
\begin{eqnarray} 
Q_\nu ^\mu \left( z \right) = \sqrt \pi  \exp \left( {i\pi \mu } \right)\frac{{\Gamma (\nu  + \mu  + 1)}}{{\Gamma (\nu  + {3 \mathord{\left/
 {\vphantom {3 2}} \right.
 \kern-\nulldelimiterspace} 2})}}{\left( {2z} \right)^{ - \nu  - 1}},\quad z \to \infty 
\end{eqnarray}
into both sides of the formula (44), we get a specification of the parameter we are looking for $
\left( {\eta, \; \tau \in \mathbb{R}} \right) $:
\begin{eqnarray} 
\cos \left( {\tau \eta } \right) = \frac{{\Gamma \left( {\nu  - i\tau  + 1} \right)\Gamma \left( {\nu  + i\tau  + 1} \right)}}{{{\Gamma ^2}\left( {\nu  + 1} \right)}}.
\end{eqnarray}
If $ \nu ,\, \tau \in \mathbb{R} $ then the equation (46) has real solutions for the parameter  $ \eta:$ in this case, according to the relation (see e. g. [14], p. 82)
\begin{eqnarray} 
\nonumber \left| {\Gamma \left( {\nu  - i\tau  + 1} \right)} \right| \le \left| {\Gamma \left( {\nu  + 1} \right)} \right|
\end{eqnarray}
one obtains from (46) an inequality $ \left| {\cos \left( {\tau \eta } \right)\,} \right| \le 1. $ Inserting the expression (46) into (44), one can write: 
\begin{eqnarray} 
& Q_\nu ^{i\tau }\left( z \right) = \exp ( - \pi \tau )\dfrac{{\Gamma \left( {\nu  + i\tau  + 1} \right)}}{{\Gamma \left( {\nu  + 1} \right)}}{Q_\nu }\left( z \right), \\ [0.2em]
\nonumber&\quad \nu,\; \tau \in \mathbb{R},\;\;\;\; \nu  > - 1,\quad \left| {\arg \left( {z - 1} \right)} \right| < \pi .
\end{eqnarray}
The functions $ {Q_\nu }\left( z \right) $ and $ Q_\nu ^{\mu }\left( z \right) $ are analytic with respect to $ \nu $  (see e. g. [15], p. 157), therefore, by analytic continuation, the relation (47) is fulfilled for $ {\mathop{\rm Re}\nolimits}\nu > - 1,\;\tau\in \mathbb{R},\;\left|{\arg\left({z-1}\right)}\right|<\pi. $ So, we have proved the next\\ 

{\bf Statement.} 
{\itshape If\; ${\mathop{\rm Re}\nolimits} \nu> - 1,\; \tau \in \mathbb{R} $ then the behavior of\; $ Q_\nu ^{i\tau }\left( z \right) $ is conditioned by the function $ {Q_\nu }\left( z \right).$}

Accordingly, one concludes that the principal value of $ Q_\nu ^{i\tau }\left( z \right) $ is analytic if $ z $ does not lie on the cut $ \left( { - \infty ,\,1} \right]. $ 

If $ \left| \nu  \right| \to \infty , $  due to the representation (see e. g. [5], 1.18(4), p. 62)
\begin{eqnarray}  
\frac{{\Gamma \left( {z + \alpha } \right)}}{{\Gamma \left( {z + \beta } \right)}} = {z^{\alpha  - \beta }},\quad \left| z \right| \to \infty , 
\end{eqnarray}
(46) gives:
\begin{eqnarray}  
\nonumber \cos \left( {\tau \eta } \right) = 1,\;\;\eta  = {{2\pi k} \mathord{\left/
 {\vphantom {{2\pi k} \tau }} \right.
 \kern-\nulldelimiterspace} \tau },\;k = 0, \pm 1,..\,\,.
\end{eqnarray}
In this case, substituting (46) into (44) and using formula (48) one can derive an asymptotic expression ($ \tau \in \mathbb{R} $): 
\begin{eqnarray}  
Q_\nu ^{i\tau }\left( z \right) = \exp ( - \pi \tau ){\nu ^{i\tau }}{Q_\nu }\left( z \right),\quad {\mathop{\rm Re}\nolimits} \nu  >  - 1,\quad \left| \nu  \right| \to \infty .
\end{eqnarray}
It is known that the function (6) has the asymptotic behavior (see e. g. [1], p. 330): 
\begin{eqnarray} 
& \;\;\;\; {Q_\nu }\left( z \right) = {\nu ^{ - {1 \mathord{\left/
 {\vphantom {1 2}} \right. 
 \kern-\nulldelimiterspace} 2}}}\exp \left[ { - \left( {\nu  + {1 \mathord{\left/
 {\vphantom {1 2}} \right.
 \kern-\nulldelimiterspace} 2}} \right)\xi } \right], \quad \xi  = {\cosh ^{ - 1}}z,  \\ [0.2em] 
& \nonumber \quad {\mathop{\rm Re}\nolimits} \nu > 0,\quad \left| \nu \right| \to \infty. 
\end{eqnarray}
The expression (50) can be rewritten as follows: 
\begin{eqnarray}  
{Q_\nu }\left( z \right) = {\nu ^{ - {1 \mathord{\left/
 {\vphantom {1 2}} \right.
 \kern-\nulldelimiterspace} 2}}}{\left( {z + \sqrt {{z^2} - 1} } \right)^{ - \nu  - {1 \mathord{\left/
 {\vphantom {1 2}} \right.
 \kern-\nulldelimiterspace} 2}}},\quad {\mathop{\rm Re}\nolimits} \nu  > 0,\quad \left| \nu  \right| \to \infty .
\end{eqnarray}
Inserting formulas (50) and (51) into (49) one obtains, respectively ($ \tau \in \mathbb{R} $):
\begin{eqnarray}  
\nonumber & Q_\nu ^{i\tau }\left( z \right) = {\nu ^{{{ - 1} \mathord{\left/
 {\vphantom {{ - 1} {2 + i\tau }}} \right.
 \kern-\nulldelimiterspace} {2 + i\tau }}}}\exp \left[ { - \pi \tau  - \left( {\nu  + {1 \mathord{\left/
 {\vphantom {1 2}} \right.
 \kern-\nulldelimiterspace} 2}} \right)\xi \,} \right],\quad {\mathop{\rm Re}\nolimits} \nu  > 0,\quad \left| \nu  \right| \to \infty , \\ [0.2em]
\nonumber & Q_\nu ^{i\tau }\left( z \right) = {\nu ^{{{ - 1} \mathord{\left/
 {\vphantom {{ - 1} {2 + i\tau }}} \right.
 \kern-\nulldelimiterspace} {2 + i\tau }}}}\exp ( - \pi \tau ){\left( {z + \sqrt {{z^2} - 1} } \right)^{ - \nu  - {1 \mathord{\left/
 {\vphantom {1 2}} \right.
 \kern-\nulldelimiterspace} 2}}},\quad {\mathop{\rm Re}\nolimits} \nu  > 0,\quad \left| \nu  \right| \to \infty .
\end{eqnarray}

In addition, inserting the asymptotic relation (see e. g. [11], p. 220)
\begin{eqnarray}
&{Q_\nu }\left( z \right) =  - \dfrac{{\ln \left( {z - 1} \right)}}{{2\Gamma \left( {\nu  + 1} \right)}} \\ 
& \nonumber \left( {z \to 1,\quad \nu  \ne  - 1,\,\; - 2,\,\;.\,\;.\,\;.\,\;,\quad \left| {\arg \left( {z - 1} \right)} \right| < \pi } \right)
\end{eqnarray}
into (47), one can find the behavior of the function $ Q_\nu ^{i\tau }\left( z \right) $ in the neighborhood of the point $ z = 1: $   
\begin{eqnarray}
& Q_\nu ^{i\tau }\left( z \right) =  - \dfrac{1}{2}\exp ( - \pi \tau )\dfrac{{\Gamma \left( {\nu  + i\tau  + 1} \right)}}{{{\Gamma ^2}\left( {\nu  + 1} \right)}}\ln \left( {z - 1} \right) \\
[0.2em]
& \nonumber \left( {z \to 1,\quad \tau \in \mathbb{R},\;\;\;\; {\mathop{\rm Re}\nolimits} \nu > - 1,\quad \;\left| {\arg \left( {z - 1} \right)} \right| < \pi } \right).
\end{eqnarray}  

According to the formulas (52) and (53) the order of singularity of the function $ Q_\nu ^\mu \left( z \right) $  at the neighborhood of the point $ z = 1 $  does not depend on the imaginary part of the parameter $ \mu . $  The character of the singularity is caused by the real part of $ \mu . $  For instance (see e. g. [5], p. 164):
\begin{eqnarray}  
Q_\nu ^\mu \left( z \right) = {1 \mathord{\left/
 {\vphantom {1 2}} \right.
 \kern-\nulldelimiterspace} 2}\exp ( - i\mu \pi \,){2^{{\mu  \mathord{\left/
 {\vphantom {\mu  {2 - 1}}} \right.
 \kern-\nulldelimiterspace} {2 - 1}}}}\Gamma \left( \mu  \right){\left( {z - 1} \right)^{ - {\mu  \mathord{\left/
 {\vphantom {\mu  2}} \right.
 \kern-\nulldelimiterspace} 2}}},\quad z \to 1,\quad {\mathop{\rm Re}\nolimits} \mu  > 0.
\end{eqnarray} \\
{\bf 4 Conclusions} \\

We have studied the Euler beta function of imaginary parameters and found its relation with the Dirac delta function [4]. For some functions, we have defined the Mellin transform of the imaginary parameters. Also, we have shown that the weak limit of the {\bf HF} function exists, for the specific set of the parameters.  	

We have found the relation with the Legendre functions of the second kind. Based on that, we have derived asymptotic formulas of the Associated Legendre function of the second kind.		

Some of the results obtained are useful in avoiding some quantum mechanical difficulties. For instance, we have calculated a nonrelativistic transition amplitude of two charged particles of continuous spectra in massive photon approximation. 		 \\        

{\bf Aknowledgment} This work was supported in part by Georgian Shota Rustaveli National Science Foundation (grant FR/417/6-100/14).

\end{document}